\documentclass[10pt]{article}
\usepackage[margin=0.85in]{geometry}
\usepackage[T1]{fontenc}
\usepackage{lmodern}
\usepackage{amsmath,amssymb,booktabs,graphicx,microtype}
\usepackage[hidelinks]{hyperref}
\usepackage{caption}
\usepackage{float}
\title{Toward Partitioned Reference Readout for Four-Level IGZO Memory: Device Statistics, Confusion Matrices, and INT4 Error Propagation}
\author{Huiyun Li\\Shenzhen University of Advanced Technology}
\date{October 2026}

\begin{document}
\maketitle

\begin{abstract}
Four-level indium--gallium--zinc oxide (IGZO) 2T0C memory needs reliable state formation, sensing references, and a path from cell errors to digital weights. We analyze 107 measured DC transfer curves from 37 isolated IGZO transistors ($W/L=3/0.5~\mu\mathrm{m}$). Four gate-bias operating points inferred from the first sweep are evaluated on later sweeps of 35 devices. Device-resolved operating points reduce second-sweep across-device current standard deviation by 49.4--52.4\%, yet retaining initial thresholds increases misclassified points from 37/140 to 55/140. Leave-one-device-out reference estimation gives 0/140 on the second sweep; freezing second-sweep thresholds yields 6/140 on the third. Sampling eight reference transistors on the third sweep gives a mean held-out error fraction of 0.15\% over 2,000 selections, versus 4.33\% with frozen thresholds on matched held-out devices. We present conditional confusion matrices, a counterfactual common-gain sensitivity, and an INT4 error-propagation proxy. A simultaneous four-state bank with eight reference cells per state requires at least 32 physical reference cells, 12.5\% of a proposed 256-cell data bank before peripheral overhead. A charge-conservation model and a partitioned hierarchical error model specify the pulse, retention, and array measurements needed to test transfer from DC device statistics to pulsed 2T0C operation.
\end{abstract}


\section{Introduction}
\subsection{Why multilevel IGZO memory needs reliable readout}
The capacity and movement of model parameters, activations, and intermediate results place demanding density and bandwidth requirements on memory close to computation. Placing more useful states in a compact, vertically integrated cell is one route to greater effective density. Amorphous indium--gallium--zinc oxide (IGZO) thin-film transistors (TFTs) are attractive here because their low off-state current can limit loss from a stored node and their low-temperature processing can support back-end-of-line integration above silicon logic~\cite{imec2021,li2025,liao2025}. A two-transistor, zero-explicit-capacitor (2T0C) gain cell uses a write transistor to charge a node associated with the read transistor and senses a resulting read current. The cell dispenses with a separately fabricated storage capacitor; it does not dispense with node capacitance or with a sensing peripheral.

Moving from a binary cell to four current levels nominally doubles stored bits per cell, but it changes the read problem. Three boundaries must remain between adjacent state-current distributions over device-to-device variation, write history, hold time, temperature, and read-path offset. A write scheme that makes each state tighter may translate the entire distribution or alter its skew. Conversely, a reference derived at one time may cease to be appropriate after the devices change. The physical spacing of four levels is therefore only one part of read reliability. The locations of the three decisions, and the cost of keeping them aligned, are equally important.

These issues are sharpened by the distinction between a cell and an array. In a cell experiment, a current may be measured with a laboratory instrument under chosen biases. In an array, the sensed quantity passes through selected word lines, bit lines, switches, bias generators, reference distribution, and a comparator or converter. Shared paths can introduce correlated errors that are invisible in isolated devices. Cell demonstrations of multibit retention or speed establish important feasibility, but do not supply a transferable array-level error rate or an energy figure for a new sensing architecture.

\subsection{Question and evidence boundary}
This study addresses a narrower, directly testable question: on the available IGZO TFT measurements, what are the separate effects of selecting device-resolved gate-bias points and updating three decision references? We use the first measured transfer sweep to infer four gate-voltage operating points, then use later sweeps as out-of-sample current observations at those points. This is an offline surrogate for possible current states, not charge programming of a 2T0C cell. We compare shared and device-resolved operating points, fixed and updated references, and the sensitivity of an update to the number of reference devices.

The work provides a traceable four-state mapping for 107 curves from 37 devices; conditional confusion matrices under fixed and updated references; a measured-data counterfactual for common read-current gain; a reference-count and minimum-area accounting; and a limited INT4 numerical-error propagation. We state separately the pulse and array measurements needed to test actual retention, shared-error, and low-energy-sensing hypotheses.

\section{Research Context and Related Work}
\subsection{International development of IGZO gain-cell memory}
Early 2T0C IGZO demonstrations established that a write/read transistor pair can retain useful information without a discrete capacitor. imec reported a capacitorless IGZO cell at IEDM 2020 and subsequently described retention, endurance, scaling, and fast operation at IEDM 2021~\cite{imec2021,belmonte2021}. The main architectural attraction is the combination of low write-transistor leakage and monolithic stacking. However, write-node loss, bias-induced threshold shifts, contact and interface variability, and coupling during read/write remain relevant even when nominal retention is long. A reported cell retention time is defined under its own voltage-window criterion and conditions; it cannot be equated with guaranteed separation of four current distributions in an independently fabricated array.

Recent work extends beyond a single binary cell. A heterogeneous capacitorless 2T0C structure has been reported with multilevel operation~\cite{lee2025}. Vertical dual-gate IGZO cells have also shown a compact $4F^2$ layout and 16 distinguishable programmed currents~\cite{liao2025}. These advances emphasize cell structure and density. They do not remove the need to specify how noisy or drifting array currents are mapped to decisions, how many references are used, or when those references are refreshed. Published figures for different channel materials, cell geometries, bias schedules, and retention criteria should be compared by evidence type rather than collapsed into a single performance ranking.

\subsection{Progress by research groups in China}
Chinese groups have advanced both the device physics and the integration of multibit oxide memory. Hu \emph{et al.} reported optimized IGZO FET-based capacitorless DRAM with three-bit operation and retention of 10 ks at room temperature and 7 ks at $85^\circ$C under their stated criterion~\cite{hu2022}. Work from the Institute of Microelectronics, Chinese Academy of Sciences, used dual-gate and in-cell threshold compensation to improve multibit state statistics~\cite{chen2023,chen2025}. Li \emph{et al.} fabricated an $8\times8$ three-dimensionally stacked IGZO 2T0C array, demonstrated three-bit states with more than 100-s retention, and used the array in a computing experiment~\cite{li2025}. Liao \emph{et al.} demonstrated self-aligned vertical dual-gate integration and four-bit cell operation~\cite{liao2025}. These are distinct device, cell, and array achievements; none is a measurement of the sensing-chain performance of the devices analyzed here.

Other studies expose why apparently clean programmed levels may change after writing. Fast $I$--$V$ experiments by Zheng \emph{et al.} separated capacitive coupling, mobile-charge sharing, and trapped-charge release in multibit ALD-IGZO 2T0C writing, identifying trapped-charge-related crosstalk as a challenge~\cite{zheng2024}. Current-based programming has also been explored to improve multibit consistency across multiple cells~\cite{mao2026}. These mechanisms motivate measurements resolved in time and across cells. Our three quasi-static sweeps cannot identify any one of them; a sweep-index effect must not be assigned a retention or trap time constant.

\subsection{Read references and the remaining gap}
Multilevel sensing maps analog observations to digital states, so its error depends on both the distributions and the location of the decision boundaries. Reference adaptation has been studied in other memory families; for example, measured relaxation in multilevel resistive memory motivated updated decisions~\cite{he2021}. That result is an analogy about decision geometry, not evidence that RRAM drift dynamics or adaptation gains transfer to IGZO. IGZO papers have primarily established cell structures, multibit state creation, retention, or array computation. The device-to-reference interaction under a controlled train/test split remains less explicit in the cited work.

Table~\ref{tab:prior} locates the present contribution by evidence scope. It is deliberately not a numerical benchmark: a single-cell retention measurement, a stacked-array computing demonstration, and an offline TFT classification answer different questions. Our hypothesis is that \emph{both} state formation and reference placement must be evaluated together, with an explicit account of when each was estimated. The present dataset lets us test that hypothesis at the device-current level. It cannot establish whether row, column, and sense-amplifier contributions are separately observable, or whether periodic versus triggered updates save energy after adding their peripheral cost.

\begin{table}[H]
\centering
\caption{Selected IGZO memory progress and the evidence gap addressed here. Reported capabilities are specific to the cited structures and conditions.}
\label{tab:prior}
\begin{tabular}{p{.18\linewidth}p{.34\linewidth}p{.38\linewidth}}
\toprule
Work & Primary experimental scope & Readout question left for this study \\
\midrule
imec~\cite{imec2021,belmonte2021} & Early IGZO 2T0C cell, retention and scaling & Four-level reference alignment across devices \\
Hu \emph{et al.}~\cite{hu2022} & Three-bit cell, speed and retention & Held-out classification after level selection \\
Chen \emph{et al.}~\cite{chen2023,chen2025} & Dual-gate, in-cell compensation & External decision-reference drift \\
Li \emph{et al.}~\cite{li2025} & $8\times8$ stacked array and computation & Separated reference-sample uncertainty \\
Liao \emph{et al.}~\cite{liao2025} & Compact vertical dual-gate cell, four-bit operation & Peripheral reference cost and read decision \\
This work & 37 isolated TFTs, repeated DC sweeps & Offline level/reference interaction only \\
\bottomrule
\end{tabular}
\end{table}

The present files contain isolated TFT $I_D$--$V_G$ sweeps. They contain neither the write/storage node of a 2T0C cell nor row/column addresses, reference cells, or sense-amplifier measurements. A separate design archive contains exploratory pre-layout CMOS sensing simulations, but not a measured IGZO array. Thus an \emph{error} below is an incorrectly classified interpolated state point, not an observed stochastic memory read error.

\section{Charge Dynamics and Identifiability}
\subsection{A physically constrained model for a future 2T0C experiment}
In a real 2T0C cell, the write transistor charges a storage node and the read transistor maps its voltage to a sensed current. After the write transistor turns off, a general charge-conservation model is
\begin{equation}
C_{\mathrm{eff}}(V_{SN})\frac{dV_{SN}}{dt}
=-I_{\mathrm{leak}}(V_{SN},T,\theta_W)
+I_{\mathrm{trap}}(z,V_{SN},T)
+\sum_k C_{SN,k}\frac{dV_k}{dt},\quad
\dot z=g(z,V_{SN},T,h),
\label{eq:charge}
\end{equation}
where $C_{\mathrm{eff}}$ includes intrinsic and wiring capacitance; $I_{\mathrm{leak}}$ is net write-off and node leakage; $I_{\mathrm{trap}}$ is a signed capture/release contribution; $V_k$ denotes coupled control lines; and $h$ records bias history. The read path obeys
\begin{equation}
I_{\mathrm{obs}}=F_{TR}(V_{SN},V_{RBL},V_{RWL},T,\theta_{TR})
+I_{\mathrm{wire}}+I_{\mathrm{sense}}.
\label{eq:read}
\end{equation}
These equations define mechanisms and measurements, not fitted time constants. The available isolated-TFT curves constrain parts of $F_{TR}$ near their measured DC bias. They do not determine $C_{\mathrm{eff}}$, the actual write-off leakage, trap occupancy, or $V_{SN}(t)$ under pulsed operation. Multiple values of these unobserved quantities can give the same transfer curves and arbitrarily different retention times. Consequently, projecting the measured DC statistics onto 1--5000~s would be a scenario conditional on chosen parameters, not a validation of 5000-s retention or bit-error rate.

The supplied design archive contains a spreadsheet-like $C_{SN}\Delta V/I_{off}$ retention screen and a 2T0C behavioral macro with assumed node capacitance and hold resistance. Neither is a pulse/hold measurement. To identify Eq.~\eqref{eq:charge}, we require actual four-state write pulses, terminal reads after undisturbed holds, separate repeated-read groups, cell capacitance or its bounds, low-current leakage with an instrument floor, and temperature and pulse-width variation. Leakage-only and leakage-plus-trap models must be checked on held-out cells and retention episodes. The maximal hold satisfying a predeclared worst-state error constraint, with uncertainty, would then define effective retention time; a target of 5000~s is not itself a result.

\subsection{Partitioned hierarchical readout model}
For array partition $q$, row $r$, column $c$, written state $s$, and sense channel $j$, we propose the observational hierarchy
\begin{equation}
Y_{qrcsj}(t)=F_s[V_{SN,qrc}(t),T,\theta_{qrc}]
+U_{\mathrm{wafer}}+U_q(t)+A_r(t)+B_c(t)
+G_q(x_{rc},t)+O_j(t)+\epsilon_{qrcsj}(t).
\label{eq:hierarchy}
\end{equation}
The terms represent wafer/batch variation, partition-shared drift, row and column effects, position-dependent wiring, sense-channel offset, and local residual error. The present CSVs have neither array coordinates nor channel labels; repeated sweeps also confound elapsed time with bias history. Thus none of the row, column, spatial, and channel terms is identifiable from them. Crossed row/column reads, same-cell repeated reads, channel swaps, address swaps, and calibrated current-source bypasses are interventions needed before physical attribution. Where terms remain confounded, the model should retain joint parameter ranges rather than assign unique mechanisms.

If data and reference drift have standard deviations $\sigma_d$ and $\sigma_r$ with correlation $\rho$, simple differencing gives residual variance $\sigma_d^2+\sigma_r^2-2\rho\sigma_d\sigma_r$. Reference tracking is therefore effective only when its cells match the data in location, state, age, bias, and read path sufficiently to make the relevant drift correlated. Shifting common thresholds cannot undo irreducible overlap of two local state distributions. For each partition, a measured conditional matrix $P(\hat s\mid s,t,T,q)$ should be used to constrain the worst state and adjacent-state errors, rather than an average error alone.

\section{Measurements and Offline Analysis}
\subsection{Raw data and sweep chronology}
Each of 37 CSV files represents one $W/L=3/0.5~\mu\mathrm{m}$ IGZO TFT. Two files contain one sweep of 81 points; 35 contain three sweeps each, giving 107 sweeps in total. The B1500A measured $I_D$ at $V_D=1~\mathrm{V}$ while sweeping $V_G$ from $3$ to $-1~\mathrm{V}$ in $0.05$-V steps. We use the columns explicitly labeled \texttt{VG} and \texttt{ID} in each \texttt{DataName} block. The laboratory confirmed that the chronological order of the three blocks in each repeated-sweep file is block 3, block 2, then block 1. This order also reproduces the independently tabulated values in the project proposal.

\subsection{Four-state map and held-out evaluation}
The target currents are $\mathbf{I}^{*}=(0.225,1.670,5.490,10.300)~\mu\mathrm{A}$ for labels 00, 01, 10, and 11. Within the measured range, we monotonize the first-sweep $V_G$--$\log_{10}I_D$ relationship and use piecewise-linear inverse interpolation to find four gate-bias operating points for each device. The mean of these voltages across all 37 devices defines a shared-operating-point control. Each device's own voltages define device-resolved operating points. Forward interpolation on the second and third sweeps gives corresponding currents. The two devices with only one sweep contribute to shared levels but not to validation. These labels do not represent actual programmed or retained charge states.

The initial references are the adjacent target-current midpoints, $\boldsymbol{\tau}_{0}=(0.9475,3.5800,7.8950)~\mu\mathrm{A}$. For each of the 35 held-out devices, we derive three second-sweep thresholds from the other 34 devices. For adjacent states $s$ and $s+1$, the threshold is
\begin{equation}
\tau_s=\tfrac12\left[\max_{i\in\mathcal{R}} I_{i,s}+\min_{i\in\mathcal{R}} I_{i,s+1}\right],
\label{eq:threshold}
\end{equation}
where $\mathcal{R}$ is the set of reference devices. The frozen-reference experiment estimates thresholds from all 35 second sweeps and applies them without change to the third sweeps. The denominator in each full-sweep error count is $35\times4=140$ device--state points. Four points from one device are correlated and should not be treated as four independent memory trials.

\subsection{Reference-sampling emulation}
On the third sweep, we randomly reserve $k\in\{1,2,4,8,16\}$ devices as references, compute the three midpoint thresholds with Eq.~\eqref{eq:threshold}, and classify only the remaining $35-k$ devices. We repeat this 2,000 times for each $k$ with a fixed random seed. We also evaluate the frozen second-sweep thresholds on the \emph{same} held-out devices in every trial. The state-point error fraction is divided by $4(35-k)$. This emulation assumes that all four labeled currents of a reference device can be obtained at the update. It excludes the area, programming, time, and energy cost of doing so.

The four currents assigned to one TFT in this offline procedure cannot be sensed simultaneously as four stored states from one real 2T0C cell. An implementable bank requires four distinct known-state cells per sampled TFT equivalent, or a sequential reprogramming procedure whose writes and delay must be counted. Repeated random choices of the same 35 devices characterize selection sensitivity within this batch; they do not create new independent devices for an array BER estimate.

\section{Results}
\subsection{Operating-point selection narrows but shifts the states}
The shared gate-bias voltages inferred from all 37 first sweeps are 0.905, 1.097, 1.277, and 1.409~V. Device-resolved operating points on the second sweep yield the measured-current statistics in Table~\ref{tab:currents}. Across-device sample standard deviation decreases by 49.4--52.4\% relative to the shared-level control. The empirical adjacent-state gaps change from $(-0.054,-0.609,-2.795)~\mu\mathrm{A}$ to $(0.650,1.938,2.020)~\mu\mathrm{A}$. The better-separated distributions shift below the original targets; a narrower distribution is therefore insufficient if the reference is left in its original location.

\begin{table}[t]
\centering
\caption{Second-sweep currents at device-resolved gate-bias points; $n=35$ devices. Standard deviations are across-device sample values.}
\label{tab:currents}
\begin{tabular}{lrrrr}
\toprule
State & 00 & 01 & 10 & 11 \\
\midrule
Mean ($\mu$A) & 0.113 & 1.061 & 4.016 & 8.052 \\
Sample SD ($\mu$A) & 0.044 & 0.251 & 0.622 & 0.965 \\
SD reduction (\%) & 52.2 & 49.4 & 50.8 & 52.4 \\
\bottomrule
\end{tabular}
\end{table}

\begin{figure}[t]
\centering\includegraphics[width=.84\linewidth]{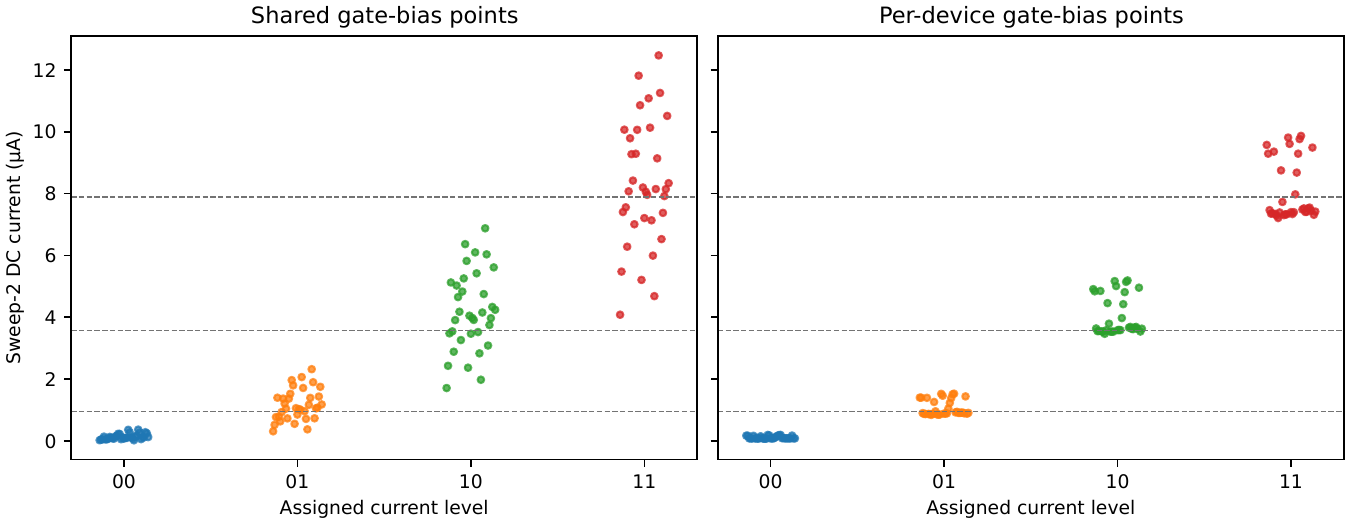}
\caption{Second-sweep four-state currents under shared and device-resolved gate-bias points. Dashed lines indicate the initial, unchanged thresholds. Each point corresponds to one device and assigned label.}
\label{fig:state}
\end{figure}

\begin{figure}[t]
\centering\includegraphics[width=.84\linewidth]{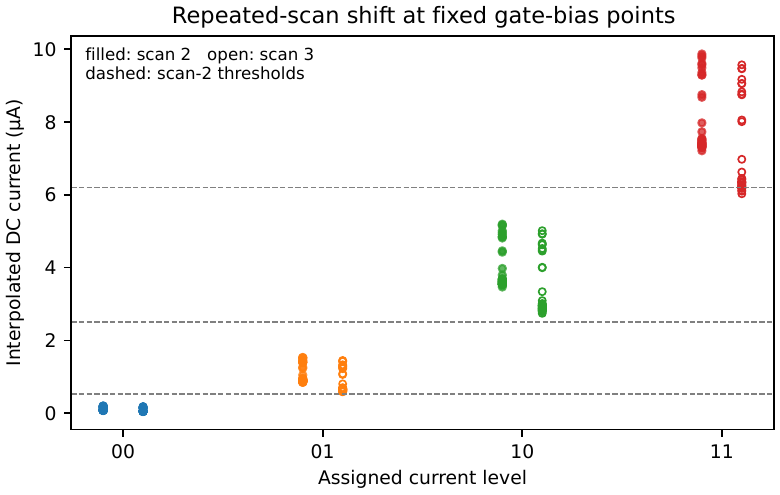}
\caption{Device-resolved currents on the second (filled markers) and third (open markers) sweeps. Dashed lines show thresholds derived from the second sweep and held fixed on the third.}
\label{fig:shift}
\end{figure}

With shared gate-bias points and initial references, 37/140 second-sweep state points are misclassified (Fig.~\ref{fig:state}). Device-resolved points with the same references give 55/140. When each test device is held out and Eq.~\eqref{eq:threshold} is applied to the other 34 devices, the mean thresholds are $(0.520,2.499,6.203)~\mu\mathrm{A}$, giving 0/140 misclassified second-sweep points. This is a finite-sample classification result, not a zero bit-error-rate estimate. The thresholds derived from all second-sweep devices and frozen for the third sweep yield 6/140 misclassified points. Each is a label-11 point read as label 10, and the six points arise from six different devices (Fig.~\ref{fig:shift}).

\subsection{Reference quantity affects robustness}
\begin{figure}[t]
\centering\includegraphics[width=.68\linewidth]{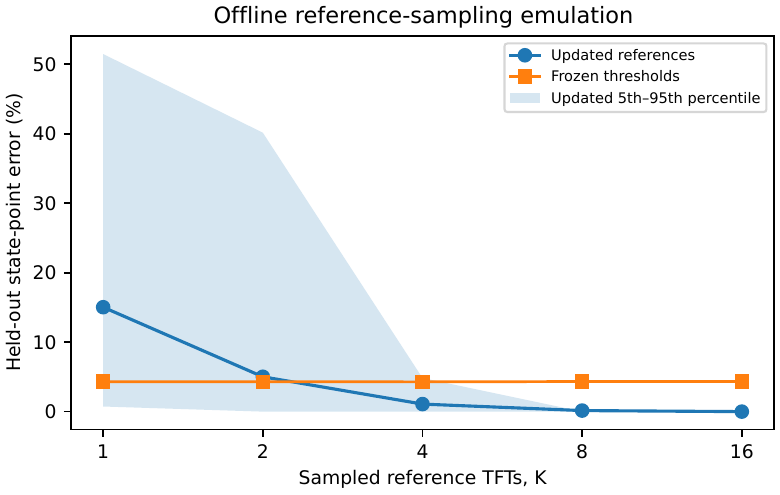}
\caption{Held-out state-point error fraction versus reference-device count in the offline emulation. The shaded band spans the 5th--95th percentiles for the updated method across random reference selections.}
\label{fig:reference}
\end{figure}

The random-reference experiment is summarized in Table~\ref{tab:reference} and Fig.~\ref{fig:reference}. One or two reference devices produce unstable thresholds and can perform worse than the frozen control. With four reference devices, the mean held-out error fraction is 1.08\%, compared with 4.28\% for the frozen thresholds evaluated on the same held-out devices. With eight references, these fractions are 0.15\% and 4.33\%, respectively. Increasing the number of references reduces the uncertainty due to reference-device selection, but consumes more devices and implies more update work in a realizable circuit. The experiment measures no such cost and therefore does not establish an energy-optimal reference count.

\begin{table}[t]
\centering
\caption{Third-sweep offline reference-sampling emulation. Entries are means over 2,000 random draws; updated and frozen methods use identical held-out devices within each draw.}
\label{tab:reference}
\begin{tabular}{rrrr}
\toprule
References $k$ & Held-out devices & Updated error (\%) & Frozen error (\%) \\
\midrule
1 & 34 & 15.03 & 4.30 \\
2 & 33 & 5.02 & 4.29 \\
4 & 31 & 1.08 & 4.28 \\
8 & 27 & 0.15 & 4.33 \\
16 & 19 & $<0.01$ & 4.33 \\
\bottomrule
\end{tabular}
\end{table}

\section{Confusion Matrices and Counterfactual Sensitivity}
\begin{figure}[t]
\centering\includegraphics[width=\linewidth]{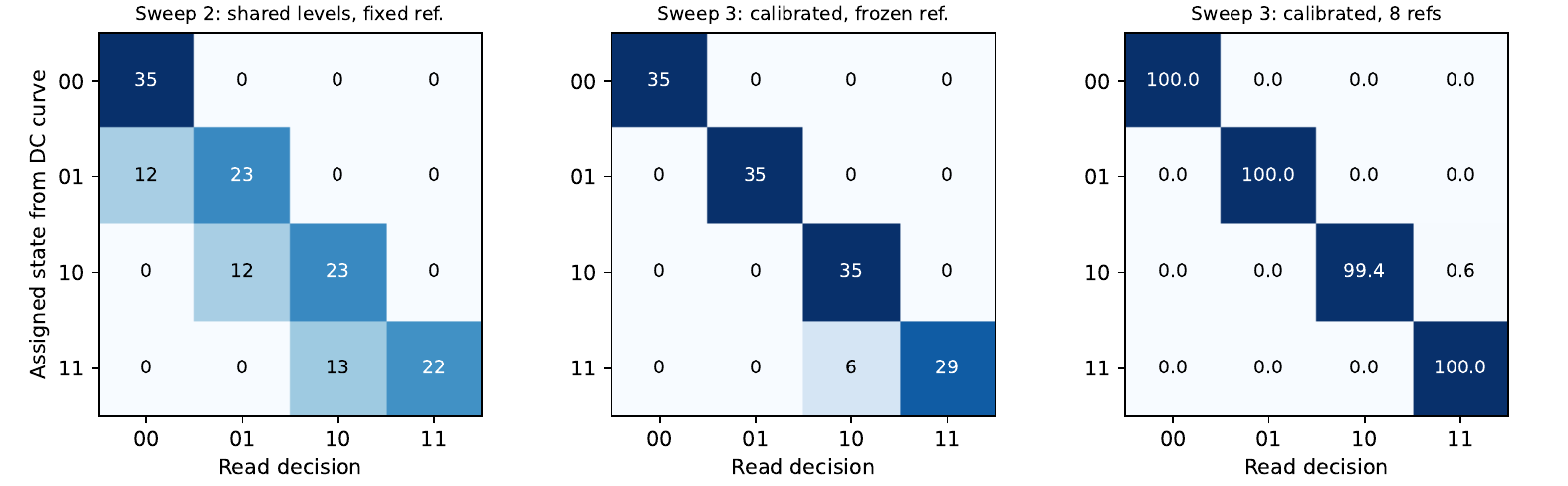}
\caption{Four-state offline confusion. Left and center: device counts from the measured DC curves. Right: row-conditional percentages from repeated $k=8$ reference selections on the same 35-device population.}
\label{fig:confusion}
\end{figure}

For assigned label $s$ and decision $h$, let $M_{sh}$ count devices classified as $h$ and $P_{sh}=M_{sh}/\sum_h M_{sh}$ denote its row-normalized conditional frequency. Figure~\ref{fig:confusion} compares an initial fixed-reference control, the third-sweep frozen-reference case, and the third-sweep $k=8$ update. The first two panels are counts of 35 devices per assigned label; the last is a frequency averaged over 2,000 overlapping reference selections, each evaluated on 27 held-out devices. They have different statistical units. In the frozen third sweep, all six errors are $11\rightarrow10$. In the sampled-reference panel, occasional $10\rightarrow11$ errors reflect threshold uncertainty from reference selection. Neither panel estimates stochastic read errors from repeated programmed 2T0C cells.

\begin{figure}[t]
\centering\includegraphics[width=.72\linewidth]{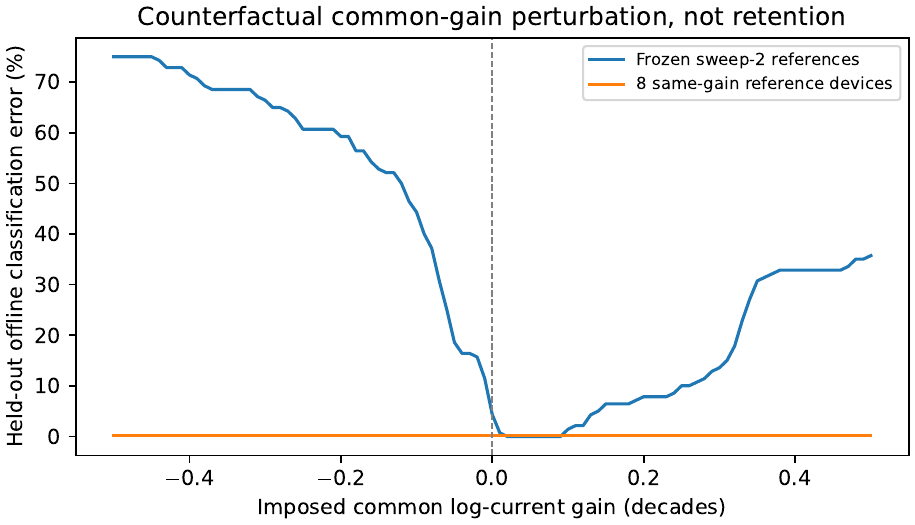}
\caption{Counterfactual shared-gain perturbation applied to measured third-sweep currents. This is a scenario, not a time-domain or retention measurement.}
\label{fig:gain}
\end{figure}

To isolate the mathematical benefit of tracking a perfectly shared current-gain term, we impose a counterfactual multiplier $10^g$ on all third-sweep test and reference currents, $g\in[-0.5,0.5]$. Frozen second-sweep thresholds remain fixed. A reference threshold recomputed on the scaled reference currents transforms as $\hat\tau_s(g)=10^g\hat\tau_s(0)$, so its decisions are invariant to this \emph{exactly shared} gain. Figure~\ref{fig:gain} displays the resulting sensitivity. The axis is log current gain, not elapsed hold time. Spatially unequal gain, reference aging mismatch, nonlinear state movement, and state overlap all break the exact cancellation. This experiment therefore makes no prediction of retention time or trap kinetics.

\section{Reference Hardware and INT4 Error Propagation}
\subsection{Minimum physical cost of sampled references}
\begin{figure}[t]
\centering\includegraphics[width=.75\linewidth]{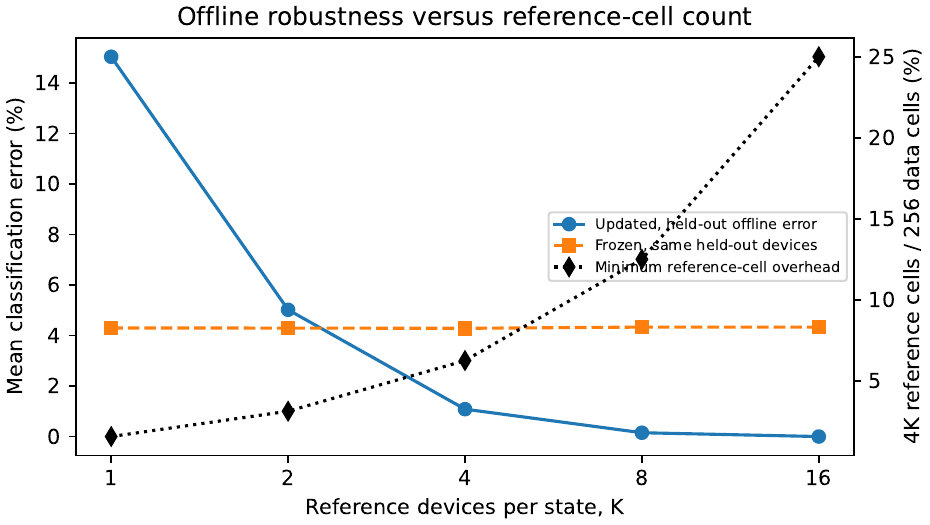}
\caption{Offline held-out error and minimum simultaneous four-state reference-cell count. The right axis is $4K/256$, a cell-count lower bound rather than physical area or energy.}
\label{fig:cost}
\end{figure}

If $K$ simultaneously available known-state cells are used for each of four states, at least $4K$ physical reference cells are needed. For a proposed $16\times16$ data array, $K=8$ requires 32 reference cells, a 12.5\% cell-count overhead relative to its 256 data cells; the fraction is 32/288 if the reference cells are added alongside the bank. This excludes three comparators, reference DAC/current mirrors, trim storage, read switches, control, wires, and any spatial duplication across partitions. Figure~\ref{fig:cost} places the measured-data sampling result beside the cell-count lower bound. The apparent drop in offline error cannot alone identify an area- or energy-optimal $K$.

For an update that reads all references, a basic accounting is
\begin{equation}
E_{\mathrm{update}}\geq4K E_{\mathrm{ref\ read}}+E_{\mathrm{threshold}}+E_{\mathrm{control}},\qquad
E_{\mathrm{amortized}}=E_{\mathrm{data\ read}}+
\frac{E_{\mathrm{update}}+E_{\mathrm{ref\ write}}}{N_{\mathrm{data\ reads}}}.
\label{eq:energy}
\end{equation}
Update latency also depends on whether four-state references are sensed in parallel, share a multiplexed path, or are reprogrammed. The present records do not provide measured values for the terms in Eq.~\eqref{eq:energy}; the equation states what a circuit comparison must count. Fixed, periodic, and risk-triggered updates should be compared at the same worst-state error target and access history.

\subsection{INT4 error propagation}
The proposed W4A8 path reconstructs one INT4 word from two 2-bit cells. Let high and low two-bit symbols be $H=2H_1+H_0$ and $L=2L_1+L_0$. Under signed two's-complement decoding,
\begin{equation}
w=-8H_1+4H_0+2L_1+L_0.
\label{eq:w4}
\end{equation}
If the conditional confusion kernel $P$ applies independently to both cells and all 16 words have equal prior probability, the expected squared numeric error follows by enumerating all $(H,L,\hat H,\hat L)$ with weight $P_{H\hat H}P_{L\hat L}/16$. For the third-sweep offline matrices, this illustration gives $0.862$ least-significant-bit (LSB) root-mean-square error for frozen references and $0.160$ LSB for the $k=8$ sampling distribution. These are deterministic propagations of a TFT-DC classification proxy under explicitly imposed priors and independence. Actual W4 weight frequencies, within-word correlated errors, scaling and W8 reconstruction, A8 activations, network architecture, and measured array errors are absent. Consequently these numbers do not imply a neural-network accuracy or perplexity change. A future hardware-in-loop study must compare the same quantized model under ideal storage and measured block-specific confusion matrices.

\subsection{Existing circuit work and its evidence level}
We located archived exploratory ngspice runs of a 3.3-V CMOS sensing path with current-mirror front ends, a p33/n33 dynamic latch, and trimmed reference mirrors. One scripted pre-layout corner/impairment summary reports 2,700/2,700 correct binary decisions; a separate integrated sense/reference/mux proxy-mismatch summary reports 2,400/2,400. Their inputs were chosen nominal currents from isolated-TFT operating points, rather than measured pulse-written 2T0C distributions. The SMIC device models are designated for HSPICE, while these runs used ngspice; cross-simulator model accuracy is unvalidated. Their randomized impairments are proxy distributions, not foundry-qualified mismatch. No final post-layout extraction, physical sensing area, energy, or worst-case delay was present. These records support circuit feasibility exploration only and cannot be combined with the offline confusion counts to claim array BER or PPA.

\section{Required Pulse, Array, and System Experiments}
Actual cell testing should first sweep write-pulse width and amplitude for four states, then measure initial current after a specified settling time. Randomized cell groups should be terminally read after holds spanning the target interval, while matched groups are repeatedly read to separate undisturbed retention from read disturbance. Every record needs cell and wafer identity, write and read bias history, hold time, temperature, instrument range, and current waveform. The $5000$-s target should be accepted only if a predeclared worst-state risk limit and its cell-clustered uncertainty are satisfied throughout the interval.

At array level, repeated reads at crossed row/column positions, channel and address swaps, and calibrated current-source bypasses are needed to fit Eq.~\eqref{eq:hierarchy}. Reference cells should be spatially and temporally paired with data. This produces measured $P(\hat s\mid s,t,T,q)$ and residual correlation versus partition size. Circuit evaluation then adds real reference reads and writes, trim operations, comparator offsets, extracted wire parasitics, and update scheduling to Eq.~\eqref{eq:energy}. Finally, two measured 2-bit decisions should reconstruct each W4 weight; fixed W4A8 model, dataset, quantization scales, and seeds are required to compare ideal-memory and hardware-in-loop task outcomes.

\section{Limitations}
The raw batch data comprise isolated-TFT DC sweeps, not pulse-programmed or retained 2T0C states. Sweep history includes bias stress and is not equivalent to a hold-time axis. Equation~\eqref{eq:charge} and its trap and coupling terms have not been fitted. The reference-sampling experiment uses already labeled state currents and resamples one population; it omits cell-to-cell array correlation, actual reference writes, and peripheral timing. Zero errors among 140 correlated device--label points do not establish zero memory BER. Even treating 35 devices as independent Bernoulli trials for \emph{any} wrong label, 0/35 has an illustrative one-sided 95\% binomial upper bound near 8.2\%, and that independence premise is unverified. The common-gain experiment is counterfactual. The INT4 proxy assumes uniform weights and two independent reads. The archived CMOS runs lack validated model translation and complete post-layout power, area, and delay. None of these evidence levels establishes a 5000-s cell-retention claim, a measured array confusion matrix, or W4A8 task accuracy.

\section{Conclusion}
The 37-device IGZO batch permits a reproducible, device-level test of reference placement: operating-point selection narrows four current distributions but can increase mistakes under stale decisions. Conditional confusion matrices and reference resampling expose which labels fail and how reference quantity affects an offline threshold. The minimum 32-cell reference-bank count at $K=8$ and an INT4 numeric-error proxy connect that result to physical and digital design questions without claiming measured array or task performance. Charge-conservation and partitioned-error equations define the measurements needed to determine actual pulse retention, spatially correlated error, and sensing cost. A stronger device or circuit publication requires those observations, not synthetic time traces extrapolated from DC data.

\section*{Data and Code Availability}
Scripts used to parse the original CSV files and generate the figures and behavioral experiments are available with the project materials. Release of the raw measurement files is subject to the data owners' decision. The counterfactual gain and INT4 proxy are labeled in the code and results. No foundry PDK files, private proposal, or third-party circuit archive are included in this manuscript's arXiv source package.

\end{document}